\documentclass[journal]{IEEEtran}
\usepackage{amsmath,amsfonts}
\usepackage{algorithmic}
\usepackage{algorithm}
\usepackage{array}
\usepackage{textcomp}
\usepackage{stfloats}
\usepackage{url}
\usepackage{verbatim}
\usepackage{graphicx}
\usepackage{cite}

\usepackage{times}
\usepackage{latexsym}

\usepackage{microtype}
\usepackage{hyperref}
\usepackage{float}
\usepackage{url}
\usepackage{amsmath,graphicx,subfigure, amssymb, bm}
\usepackage{booktabs,arydshln,multirow}

\usepackage{ulem}
\usepackage[textsize=tiny]{todonotes}

\newcommand{\changes}[1]{{\color{black}#1}}
\newcommand{\lastchanges}[1]{{\color{black}#1}}

\begin{document}

\title{Cross-Utterance Conditioned VAE \\ for Speech Generation}

\author{
Yang Li$^\dagger$;\thanks{Yang Li and Wei Pan work with Department of Computer Science, the University of Manchester, Manchester, UK.
(e-mail: yang.li-4@manchester.ac.uk, wei.pan@manchester.ac.uk)
}
Cheng Yu$^\dagger$;\thanks{
Cheng Yu, Weiqin Zu, and Zheng Tian are with the School of Creativity and Art, ShanghaiTech University, Shanghai, China. (e-mail: yucheng@shanghaitech.edu.cn,  zuwq2022@shanghaitech.edu.cn, tianzheng@shanghaitech.edu.cn) 
}
Guangzhi Sun, Student Member, IEEE; \thanks{
Guangzhi Sun is with the Machine Intelligence Lab, University of Cambridge, Cambridge, UK. (e-mail: gs534@cam.ac.uk)
}
Weiqin Zu;
Zheng Tian;
Ying Wen, Member, IEEE;\thanks{
Ying Wen is with the School of Electronic, Information and Electrical Engineering (SEIEE), Shanghai Jiao Tong University, Shanghai, China. (e-mail: 
ying.wen@sjtu.edu.cn)
}
Wei Pan, Member, IEEE;
Chao Zhang, Member, IEEE;\thanks{
Chao Zhang is with the Department of Electronic Engineering, Tsinghua University, Beijing, China, and Department of Speech Hearing and Phonetic Sciences, UCL, London, UK. 
(e-mail: cz277@tsinghua.edu.cn)
}
Jun Wang; \thanks{
Jun Wang is with the Department of Computer Science, University College London, London, UK. (e-mail:jun.wang@cs.ucl.ac.uk)
}
Yang Yang, Fellow, IEEE; \thanks{
Yang Yang is with the Thrust of Internet of Things, The Hong Kong University of Science and Technology (Guangzhou), Guangzhou, 511453, China. (IEEE Account ID: 40335592e-mail:yyiot@hkust-gz.edu.cn, dr.yangyang@ieee.org)
}
Fanglei Sun$^\star$;\thanks{
Fanglei Sun is with the Department of Computer Science and Engineering at University of Shanghai for Science and Technology. (e-mail: sunfanglei@usst.edu.cn)
}
\thanks{
$\dagger$: Those authors are equal contribution. $\star$: Fanglei Sun is the corresponding author.
}
}
\markboth{Journal of \LaTeX\ Class Files,~Vol.~14, No.~8, August~2021}%
{Shell \MakeLowercase{\textit{et al.}}: A Sample Article Using IEEEtran.cls for IEEE Journals}

\maketitle

\begin{abstract}
Speech synthesis systems powered by neural networks hold promise for multimedia production, but frequently face issues with producing expressive speech and seamless editing. In response, we present the Cross-Utterance Conditioned Variational Autoencoder speech synthesis (CUC-VAE S2) framework to enhance prosody and ensure natural speech generation.
This framework leverages the powerful representational capabilities of pre-trained language models and the re-expression abilities of variational autoencoders (VAEs). 
The core component of the CUC-VAE S2 framework is the cross-utterance CVAE, which extracts acoustic, speaker, and textual features from surrounding sentences to generate context-sensitive prosodic features, more accurately emulating human prosody generation. 
We further propose two practical algorithms tailored for distinct speech synthesis applications: CUC-VAE TTS for text-to-speech and CUC-VAE SE for speech editing. 
The CUC-VAE TTS is a direct application of the framework, designed to generate audio with contextual prosody derived from surrounding texts. 
On the other hand, the CUC-VAE SE algorithm leverages real mel spectrogram sampling conditioned on contextual information, producing audio that closely mirrors real sound and thereby facilitating flexible speech editing based on text such as deletion, insertion, and replacement.
Experimental results on the LibriTTS datasets demonstrate that our proposed models significantly enhance speech synthesis and editing, producing more natural and expressive speech.

\end{abstract}


\section{Introduction}
\IEEEPARstart{T}{he} advent of neural network speech systems has seen widespread adoption across diverse sectors, including social media, gaming, and film production, particularly in the realm of dubbing. 
These sectors often require frequent modifications to textual content to cater to user preferences, leading to an increased demand for speech generation systems that deliver high levels of naturalness and expressiveness.

Recent advancements in Text-to-Speech (TTS) systems have leveraged their powerful expressive potential to synthesize missing or inserted words based on text transcription and original audio. These systems explicitly utilize either style tokens or variational autoencoders (VAEs)~\cite{vae, hsu2018hierarchical} to encapsulate prosody information into latent representations.
Fine-grained prosody modeling and control have been achieved by extracting prosody features at the phoneme or word-level~\cite{lee2019finegrained,sun2020generating,sun2020finegrainedvae}. However, VAE-based TTS systems often lack control over the latent space, as the sampling during inference is performed from a standard Gaussian prior. To address this, recent research has employed a conditional VAE (CVAE)~\cite{cvae} to synthesize speech from a conditional prior.

Unlike TTS, which demands extensive high-quality training data, text-based speech editing (SE) capitalizes on raw audio during inference, promoting efficient speech generation at lowered costs. 
However, most existing speech editing methods, such as VoCo~\cite{VoCo}, EditSpeech~\cite{EditSpeech}, CampNet~\cite{CampNet}, and A$^3$T\cite{a3t}, heavily rely on partially inferred TTS systems. 
These methods might cause discontinuities at editing junctions and challenges in capturing shifts in tone and prosody. In particular, modifying the transcript of a speech recording can impact the tone and prosody of the adjacent audio. The audio produced by such methods might not seamlessly integrate with the surrounding sounds, leading to an unnatural feel.
This limitation restricts the prosody and style continuity of the synthesized speech. Prosody modeling, therefore, plays an indispensable role in speech editing. 
Numerous studies have leveraged pre-trained language models (LM) to predict prosodic attributes from utterances or segments~\cite{hayashi2019pre,Kenter2020RNNbert,Jia2021PnGBERT,Futamata2021preJapan,cong2021controllable}, and have employed style tokens or Variational Autoencoders (VAEs) to encapsulate prosody as latent representations~\cite{lee2019finegrained,sun2020generating,sun2020finegrainedvae,2019Dahmaniconditional,Karanasou2021learned}.

In this work, our primary focus is on enhancing the expressiveness and naturalness of speech synthesis. To this end, we propose the Cross-Utterance Conditional Variational Autoencoder Speech Synthesis Framework (CUC-VAE S2), which significantly combines and extends our previous conference paper~\cite{cucvae, yucross}. 
This framework is designed to more accurately emulate human prosody generation by extracting acoustic, speaker, and textual features from surrounding sentences to produce context-sensitive prosodic features. Central to its architecture are the Cross-Utterance Embedding (CU-Embedding), the Cross-Utterance Enhanced CVAE, and a vocoder. 
The CU-Embedding generates phoneme-level embeddings by utilizing multi-head attention layers and taking BERT sentence embeddings from surrounding utterances as inputs. Conversely, the CUC-VAE estimates the posterior of latent prosody features for each phoneme, enhances the VAE encoder with an utterance-specific prior, and samples latent prosody features from the derived utterance-specific prior during inference.
To tackle real-world speech synthesis, we introduce two specialized algorithms for specific applications: CUC-VAE TTS for text-to-speech and CUC-VAE SE for speech editing.
The CUC-VAE TTS algorithm is a direct implementation of our framework, aiming to produce audio that carries contextual prosody using surrounding text data.
The CUC-VAE SE samples real mel spectrograms using contextual data, creating authentic-sounding audio that supports versatile text edits. We use a variational autoencoder with masked training to ensure output fidelity.


Empirically, we implemented a comprehensive set of experiments on the LibriTTS English audiobook dataset to ascertain the efficacy of our proposed CUC-VAE TTS and CUC-VAE SE systems. 
These evaluations encompassed subjective naturalness ratings, objective reconstruction metrics, and prosody diversity experiments.
Experimental results suggest that CUC-VAE TTS achieves superior performance in terms of prosody diversity, while simultaneously augmenting naturalness and intelligibility. Additionally, through subjective evaluations and a rigorous significance analysis of naturalness and similarity, it was discovered that, relative to the baseline system, CUC-VAE SE demonstrates a commendable ability to sustain high fidelity and markedly enhance naturalness across a spectrum of editing operations.

We make the following contributions:
\begin{itemize}
  \item We introduce the innovative Cross-Utterance Conditional Variational Autoencoder Speech Synthesis (CUC-VAE S2) framework to boost the expressiveness and naturalness of synthesized speech by generating context-sensitive prosodic features.
  \item We propose two practical algorithms, CUC-VAE TTS for text-to-speech and CUC-VAE SE for speech editing, both integrated with FastSpeech 2, to address real-world speech synthesis challenges.
  \item Empirical validations of the proposed CUC-VAE TTS and CUC-VAE SE systems are performed on the LibriTTS English audiobook dataset, demonstrating their effectiveness in improving prosody diversity, naturalness, and intelligibility of synthesized speech.
\end{itemize}

The rest of this paper is organized as follows: Section~\ref{sec: background} provides an overview of the relevant literature. Our proposed CUC-VAE S2 framework is detailed in Section~\ref{sec: model}, followed by the introduction of two practical algorithms in Section~\ref{sec: algo}. The experimental setup and results are discussed in Section~\ref{sec:exp_results}, and finally, Section~\ref{sec: conclusion} concludes the paper.

\section{Related Work}
\label{sec: background}
\subsection{Non-Autoregressive TTS and FastSpeech}
\label{sec: background_NARTTS}
Significant advancements have been made in non-autoregressive (Non-AR) text-to-speech (TTS) systems with regard to efficiency and fidelity, thanks to the progress made in deep learning. A Non-AR TTS system is designed to map input text sequences to acoustic features or waveforms without utilizing autoregressive decomposition of output probabilities.
Certain Non-AR TTS systems require distillation from an autoregressive model, such as FastSpeech~\cite{ren2019fastspeech} and ParaNet~\cite{peng2019paralltts}. However, recent efforts have been devoted to developing non-distillation-based Non-AR TTS systems, including FastPitch~\cite{Lancucki2021FastPitch}, AlignTTS~\cite{Zeng2020AlignTTS}, and FastSpeech 2~\cite{ren2020fastspeech}. 
\changes{Additionally, Non-AR Tacotron-based TTS systems~\cite{Elias2021ParallelTacotron, Miao2020EfficientTTS}, flow-based TTS systems~\cite{Miao2020FlowTTS, Kim2020GlowTTS}, and diffusion model TTS~\cite{Popov2021GradTTSAD, Jeong2021DiffTTSAD} have made remarkable strides in enhancing the quality and speed of speech synthesis.}

\changes{FastSpeech~\cite{ren2019fastspeech} and its follow-up works~\cite{ren2020fastspeech, Ren2022Prosospeech} employ a Non-AR sequence-to-sequence encoder-decoder model that utilizes Transformer architecture.}
The length regulator proposed by FastSpeech predicts the duration of each phoneme and then up-samples the encoded phoneme sequence based on the predicted duration. The decoder then maps the up-sampled sequence to the output acoustic feature sequence of the same length, where the calculation of each output is carried out in parallel. The absence of dependence on the previous time step output accelerates model inference. FastSpeech's length regulator is trained by distilling knowledge from an autoregressive model, providing reasonable control over the speech's rhythm.

FastSpeech 2~\cite{ren2020fastspeech} replaces the knowledge distillation for the length regulator with mean-squared error training based on duration labels. These labels are obtained from frame-to-phoneme alignment, simplifying the training process. Furthermore, FastSpeech 2 predicts pitch and energy information from the encoder output, supervised with pitch contours and L2-norm of amplitudes as labels, respectively. This additional pitch and energy prediction injects extra prosody information into the decoder, improving the naturalness and expressiveness of the synthesized speech.
\changes{Nevertheless, both FastSpeech methods and these non-autoregressive TTS systems suffer from a limitation in effectively harnessing contextual information to enhance prosody and ensure natural speech generation. 
In this paper, to address this limitation, we introduce our proposed CUC-VAE TTS method, which goes beyond by generating audio with contextual prosody derived from surrounding texts.}

\subsection{Text-based Speech Editing}
Text-based speech editing systems have emerged as powerful tools for user interaction with speech anchored on the provided transcript. Unlike TTS which necessitates extensive high-quality training data, text-based SE leverages raw audio during the inference phase. This not only enhances efficiency but also fosters flexibility in speech generation, often at a diminished cost. One of the pioneering efforts in this domain, VoCo~\cite{VoCo}, synthesizes a word via a TTS voice resembling the target speaker and subsequently refines it using a voice conversion (VC) model to ensure a match with the desired speaker's tonality.

Furthermore, there have been innovative strides in this sphere. For example, CampNet~\cite{CampNet} has introduced a context-aware Transformer optimized for this application, and A$^3$T~\cite{a3t} brought forward an alignment-aware Conformer. These systems excel in the intricate task of text-based speech editing through meticulous engineering. EditSpeech~\cite{EditSpeech} presents a unique method incorporating partial inference combined with bidirectional fusion, ensuring seamless transitions at editing junctions. Notably, VALL-E~\cite{VALl-E} showcased impressive voice cloning outcomes on expansive datasets by favoring codec representation over traditional mel spectrograms, which synergizes perfectly with language models.

Nevertheless, certain prevalent challenges persist. Many preceding methods that rely on directly copying the mel-spectrogram or waveform of untouched regions sometimes falter, resulting in disrupted prosody around the editing boundaries. They also occasionally suffer from limited adaptability to evolving contexts. Addressing these nuances, our CUC-VAE SE system harnesses the capabilities of the CUC-VAE. It meticulously models prosody by considering both adjacent utterances and the neighboring waveform in the editing vicinity, enhancing both the quality and expressiveness of the edited speech.

\subsection{VAEs in TTS}
\label{sec: background_VAE}
Variational Autoencoders (VAEs) have emerged as a popular tool in the development of TTS systems, owing to their ability to provide an explicit model for prosody variation. VAEs operate by mapping input features onto vectors in a low-dimensional latent space, wherein they capture the variability inherent in the data, before reconstructing the input using these vectors. 

Many studies have employed VAEs to capture and disentangle diverse forms of data variations in the latent space. 
For instance, Hsu et al. \cite{hsu2017learning, hsu2017unsupervised} utilized VAEs to disentangle speaker and phoneme information. 
Similarly, Akuzawa et al. \cite{akuzawa2018expressive} leveraged VAEs to model speaking style for expressive speech synthesis. 
Moreover, Hsu et al. \cite{hsu2018hierarchical, hsu2018disentangle} delved into disentangling prosody variation and speaker information using VAEs coupled with adversarial training. 
More recently, fine-grained VAEs have been employed to model prosody in the latent space for each phoneme or word \cite{sun2020generating, sun2020finegrainedvae}. 
Furthermore, vector-quantized VAEs have been applied to discrete duration modeling \cite{yasuda2021end}.
\changes{These approaches directly sample from a standard Gaussian distribution during inference period, which can introduce diversity in prosody. However, such diversity may seem random. In this paper, we integrate a conditional variational autoencoder (CVAE) into the audio generation system to refine prosody using the predicted conditional prior distribution derived from cross-utterance information.}

\subsection{Pre-trained Representation in TTS}
\label{sec: background_BERT}
It is believed that language information in both current and surrounding utterances can be utilized to infer prosody \cite{hayashi2019pre,xu2021improving,shen2018natural,fang2019towards,zhang2020unified,zhou2021dependency}. Such information is often embedded in vector representations obtained from a pre-trained language model (LM), such as BERT \cite{devlin2018bert}. Prior research has incorporated BERT embeddings at the word or sub-word level into auto-regressive TTS models~\cite{shen2018natural,fang2019towards}. More recently, Xu et al.~\cite{xu2021improving} employed chunked and paired sentence patterns from BERT.
Some studies have combined BERT with other techniques to enhance the model's expressiveness. In~\cite{zhang2020unified}, BERT was combined with a multi-task learning approach to disambiguate polyphonic characters in Mandarin's grapheme-to-phoneme (G2P) conversion. Moreover, Zhou et al.~\cite{zhou2021dependency} used a relational gated graph network with pre-trained BERT embeddings as node inputs to extract word-level semantic representations, enhancing the expressiveness of the model.
\changes{
Our proposed CUC-VAE S2 framework integrates a pretrained language model into the CVAE to extract contextual information from neighboring utterances, enhancing prosody and improving the naturalness of the generated audio.}

\begin{figure*}[ht!]
\centering
\includegraphics[width=0.8\linewidth]{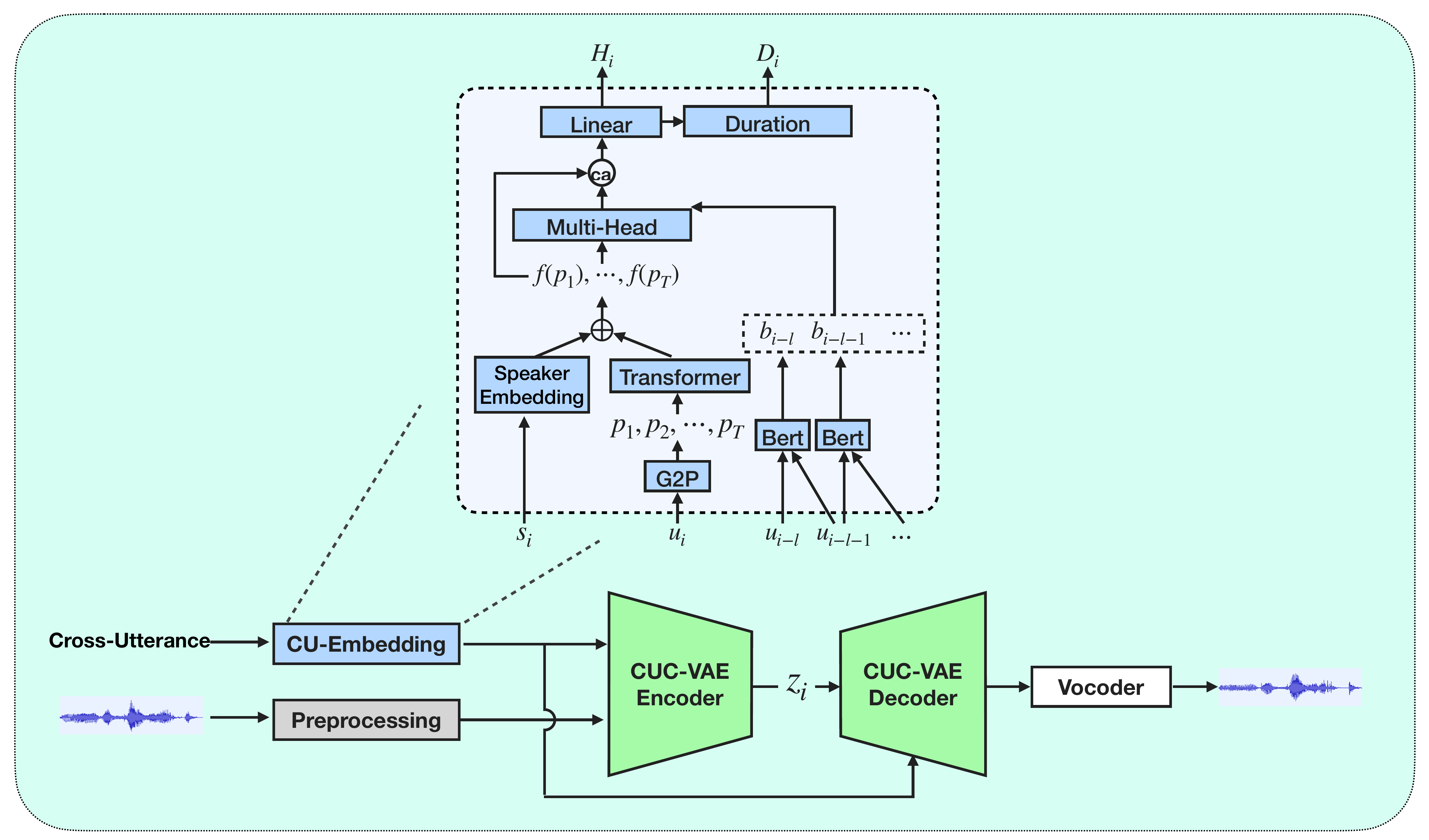}
\caption{An overview of the Cross-Utterance Conditioned VAE Speech Synthesis (CUC-VAE S2) Framework architecture. The primary CUC-VAE synthesizer utilizes textual information derived from neighboring text via the Cross-Utterance (CU) embedding, as well as audio information processed by the pre-processing module. A supplementary vocoder is incorporated with the purpose of transforming the synthesized mel-spectrogram into waveform.
}
\label{fig:model}
\end{figure*}

\section{Cross-Utterance Conditioned VAE Speech Synthesis Framework}
\label{sec: model}
\subsection{Overview}
As illustrated in Fig.~\ref{fig:model}, our research presents a sophisticated prosody speech synthesis framework. This framework distinctively integrates a conditional VAE module, designed to efficiently extract prosody from neighboring utterances. We refer to this as the Cross-Utterance Conditioned VAE Speech Synthesis Framework (CUC-VAE S2).

As illustrated in the figure, the CUC-VAE S2 primarily utilizes the CUC-VAE to model prosody extracted from neighboring utterances. Initially, we propose a Cross-Utterance (CU) embedding module to glean prosodic information from the current phoneme sentences, speaker data, and cross-utterance information. This embedding vector with predicted duration by CU-embedding denoted as $\bm H$  and $\bm D$, is subsequently forwarded to the CUC-VAE, in conjunction with pre-processed audio, to generate the target mel-spectrogram.

More specifically, the encoder $\mathcal{E}$ approximates the posterior $\bm z$ conditioned on an utterance-specific conditional prior, $\bm z_p$. This variable, $\bm z_p$, is sampled from the utterance-specific prior that has been learned from the outputs $\bm H$  and $\bm D$ of the CU-embedding module. The sampling is performed from the estimated prior by the utterance-specific prior module and is reparameterized as follows:
\begin{equation}
\label{eq:sample_z}
\bm{z} = \bm \mu \oplus \bm \sigma \otimes \bm{z}_{p},
\end{equation}
where the element-wise addition and multiplication operations are represented by $\oplus$ and $\otimes$, $\bm \mu$ and $\bm \sigma$ are predicted means and the covariance matrices from pre-processed audio information. 
The decoder $\mathcal{D}$ employed within the CUC-VAE is derived from FastSpeech 2~\cite{ren2020fastspeech}. Following the CUC-VAE synthesizer, an additional vocoder, HifiGAN~\cite{hifigan}, is implemented to transform the synthesized mel-spectrogram into an audible waveform. The vocoder can either directly utilize a pre-trained model or be fine-tuned according to the data predicted by the CUC-VAE synthesizer.  Algorithm~\ref{alg:model} provides the pseudocode for the training process of the CUC-VAE S2 framework at a high level.

Building upon the CUC-VAE S2 framework, we further propose two practical algorithms: CUC-VAE TTS for text-to-speech tasks, and CUC-VAE SE for speech editing tasks. These will be elaborated upon in the subsequent sections. The pre-processing module and CUC-VAE encoder vary across different tasks. Therefore, we will provide a detailed introduction to the CU-Embedding module in this section, while the other proposed modules will be discussed in the following sections for specific practical algorithms. 
Besides, we introduce the optimization objective in Section~\ref{sec:LO}.

\begin{algorithm}[t!]
\caption{CUC-VAE S2 Training Pseudocode}
\renewcommand{\algorithmicrequire}{\textbf{Input:}}
\renewcommand{\algorithmicensure}{\textbf{Output:}}
\label{alg:model}
\begin{algorithmic}[1]
 \REQUIRE Dataset $\mathcal{D}$
\WHILE{Not Converge}
    \STATE \textcolor{gray}{\# Speaker ID $s$, Current Utterance $u$, Cross-utterance pairs $\bm C$ and target audio $x^{target}$.}
    \FOR{$(s, u, \bm C, \changes{x^{target}}) \in \mathcal{D}$} 
     \STATE $x \leftarrow \operatorname{preprocessing}(y)$
     \STATE $\bm H, \bm D \leftarrow \operatorname{CU-Embedding}(s, u, \bm C)$ by Eq.~\ref{eq:CU_eq1} to \ref{eq:proj}
      \STATE $\bm{\mu}, \bm{\sigma}, \bm{z}_{p} \leftarrow \mathcal{E}(x, \bm H, \bm D)$
      \STATE $\bm{z} = \bm \mu \oplus \bm \sigma \otimes \bm{z}_{p}$
      \STATE $\bm \hat{x} \leftarrow \mathcal{D}(\bm{z}, \bm H, \bm D)$
     \STATE Calculate $\mathcal{L}$ (like  Eq.~\ref{eq:elbo_objective} and Eq.~\ref{obj:se}), and update parameters 
    \ENDFOR
\ENDWHILE

\end{algorithmic}
\end{algorithm}

\subsection{CU-Embedding Module}
The CU-embedding component of our proposed system incorporates cross-utterance information, phoneme sequence, and speaker information to generate a sequence of mixture encodings. 
As depicted in Fig.\ref{fig:model}, the text input consists of speaker information, the current utterance $\bm{u}_i$, and the $L$ utterances before and after the current one. The extra G2P conversion step is first performed to convert the current utterance into a phoneme sequence denoted as $\bm{P}_i=[p_1,p_2,\cdots,p_T]$, where $T$ is the number of phonemes. Additionally, the start and end times of each phoneme can be extracted using Montreal forced alignment~\cite{mfa}.

A Transformer encoder is then utilized for encoding the phoneme sequence into a sequence of phoneme encodings. Furthermore, the speaker information is encoded into a speaker embedding $\bm{s}_i$, which is added to each phoneme encoding to produce the mixture encodings $\bm{F}_i$ of the phoneme sequence.
\begin{equation}
\bm{F}_i = [\bm{f}_i(p_1),\bm{f}_i(p_2),\cdots,\bm{f}_i(p_T)],
\label{eq:CU_eq1}
\end{equation}
The vector $\bm{f}$ denotes the resulting vector obtained by adding each phoneme encoding and the speaker embedding.

To enhance the naturalness and expressiveness of the generated audio, contextual information is captured by incorporating cross-utterance BERT embeddings and a multi-head attention layer. Specifically, $2l$ cross-utterance pairs represented as $\bm{C}_i$ are extracted from $2l+1$ adjacent utterances $[\bm{u}_{i-l},\cdots,\bm{u}_i,\cdots,\bm{u}_{i+l}]$ using Equation (\ref{eq:uttpairs}).

\begin{small}
\begin{equation}
\resizebox{\linewidth}{!}{
$
    \bm{C}_i = [c(\bm{u}_{i-l}, \bm{u}_{i-l+1}),\cdots, c(\bm{u}_{i-1}, \bm{u}_{i}),\cdots, c(\bm{u}_{i+l-1}, \bm{u}_{i+l})],
$
    }
    \label{eq:uttpairs}
\end{equation}
\end{small}
\changes{The cross-utterance pairs, represented as $c(u_k,u_{k+1})= \{[\text{CLS}], \bm{u}_k, [\text{SEP}], \bm{u}_{k+1}\}$, comprise adjacent utterances $\bm{u}_k$ and $\bm{u}_{k+1}$. The [CLS] token is prefixed at the start of each pair, while the [SEP] token is inserted at the boundary of each sentence for BERT tracking purposes.
The decision to utilize the [CLS] token embedding as the representation for cross-utterance pair embedding stems from its capacity to capture the semantic essence of the entire sentence. Leveraging this representation, we aim to distill crucial semantic information from the utterance pair, facilitating downstream tasks effectively.}
Subsequently, the $2l$ cross-utterance pairs are fed into BERT to capture cross-utterance information. The resulting output consists of $2l$ BERT embedding vectors, each of which is obtained by taking the output vector at the position of the [CLS] token and projecting it to a 768-dimensional vector for each cross-utterance pair, as demonstrated below:
$$
\bm{B}_i = [\bm b_{-l}, \bm b_{-l+1}, \cdots, \bm b_{l-1}],
$$
where each vector $\bm b_k$ in $\bm B_i$ represents the BERT embedding of the cross-utterance pair $c(\bm u_k,\bm u_{k+1})$. To extract CU-embedding vectors for each phoneme specifically, a multi-head attention layer has been incorporated to merge the $2l$ BERT embeddings into a single vector, as demonstrated in Equation~\eqref{eq:mha}.
\begin{equation}
    \bm{G}_{i}=\text {MHA}(\bm{F_i} \bm{W}^{\text{Q}}, \bm{B}_i \bm{W}^{\text{K}}, \bm{B}_i \bm{W}^{\text{V}}),
    \label{eq:mha}
\end{equation}
The multi-head attention layer is denoted as MHA$(\cdot)$, with $\bm{W}^{\text{Q}}$, $\bm{W}^{\text{K}}$, and $\bm{W}^{\text{V}}$ serving as linear projection matrices, and $\bm{F_i}$ representing the sequence of mixture encodings for the current utterance, which functions as the query in the attention mechanism. 
To simplify the notation, the expression in Equation~\eqref{eq:mha} is denoted as $\bm{G}_{i}=[\bm{g}_{1}, \bm{g}_{2},\cdots,\bm{g}_{T}]$, where the length of the multi-head attention mechanism is $T$, and each element is concatenated with its corresponding mixture encoding. 
Subsequently, these concatenated vectors are projected by another linear layer to generate the final output $\bm{H}_i$ of the CU-embedding, denoted as $\bm{H}_i=[\bm{h}_1,\bm{h}_2,\cdots,\bm{h}_T]$, as illustrated in Equation~\eqref{eq:proj}.
\begin{equation}
    \bm{h}_{t} = [\bm{g}_{t},\bm{f}(p_t)]\bm{W},
    \label{eq:proj}
\end{equation}
where $\bm{W}$ is a linear projection matrix.
Furthermore, an extra duration predictor takes $\bm{H}_i$ as its input and predicts the duration $\bm D_i$ for each phoneme.

\subsection{Learning Objective}
\label{sec:LO}
In accordance with the Evidence Lower Bound (ELBO) objective function~\cite{vae}, we can derive our ELBO objective, which is expressed as follows:
\begin{equation}
\label{eq:elbo_objective}
\resizebox{\linewidth}{!}{
$
\begin{aligned}
\mathcal{L}(\bm x\mid{\bm H, \bm{D}})&=\mathbb{E}_{q_{\phi}(\bm{z} \mid{\bm D, \bm H})} [\log p_\theta(\bm x \mid \bm z, \bm D, \bm H)]\\
&-\beta_1 \sum_{n=1}^{t} D_{\mathrm{KL}}\left(q_{\phi_1}\left({\bm z^n} \mid{\bm z_{p}^n, \bm x}\right) \| q_{\phi_2} \left({\bm z_{p}^n} \mid{\bm D, \bm H}\right)\right)\\
&-\beta_2 \sum_{n=1}^{t} D_{\mathrm{KL}}\left(q_{\phi_2}\left({\bm z_{p}^n} \mid{\bm D, \bm H}\right) \| p(\bm z_{p}^n)\right).
\end{aligned}
$
}
\end{equation}
The index $i$, which signifies the current instance, has been omitted for the sake of simplicity. 
Here, $\theta$ denotes the parameters of the decoder module, 
whereas $\phi_1$ and $\phi_2$ represent two components of the mask CUC-VAE encoder $\phi$ that derive $\bm z$ from $\bm z_{p}, \bm x$ and $\bm z_{p}$ from $\bm D,\bm H$, respectively. 
Additionally, $\beta_1$ and $\beta_2$ represent two balancing constants, and $p(\bm z_{p}^n)$ is selected to be a standard Gaussian distribution, i.e., $\mathcal{N}(0,1)$.
Moreover, $\bm z^n$ and $\bm z_{p}^n$ denote the latent representation for the $n$-th phoneme, and $t$ corresponds to the length of the phoneme sequence.

\section{Practical Algorithms}
\label{sec: algo}
In this section, we will introduce two practical algorithms based on CUC-VAE S2 framework, named CUC-VAE TTS for text-to-speech task and CUC-VAE SE for speech editing task.
As described in Section~\ref{sec: model}, CUC-VAE S2 framework mainly proposes a variational autoencoder conditioned on contextual embedding to achieve the ultimate goal of prosody speech synthesis for different tasks.
For TTS tasks, we propose the CUC-VAE TTS algorithm to produce speech with natural and expressive prosody.
Besides, the CUC-VAE SE algorithm is designed to efficiently edit speech without resorting to the splicing of the generated and original audio.
The details of two algorithms are given in Section~\ref{sec:CUC_VAE_TTS} and Section~\ref{sec:CUC_VAE_SE}.

\subsection{CUC-VAE TTS Algorithm for Text-to-Speech}
\label{sec:CUC_VAE_TTS}

\begin{figure}
\centering
\includegraphics[width=0.9\linewidth]{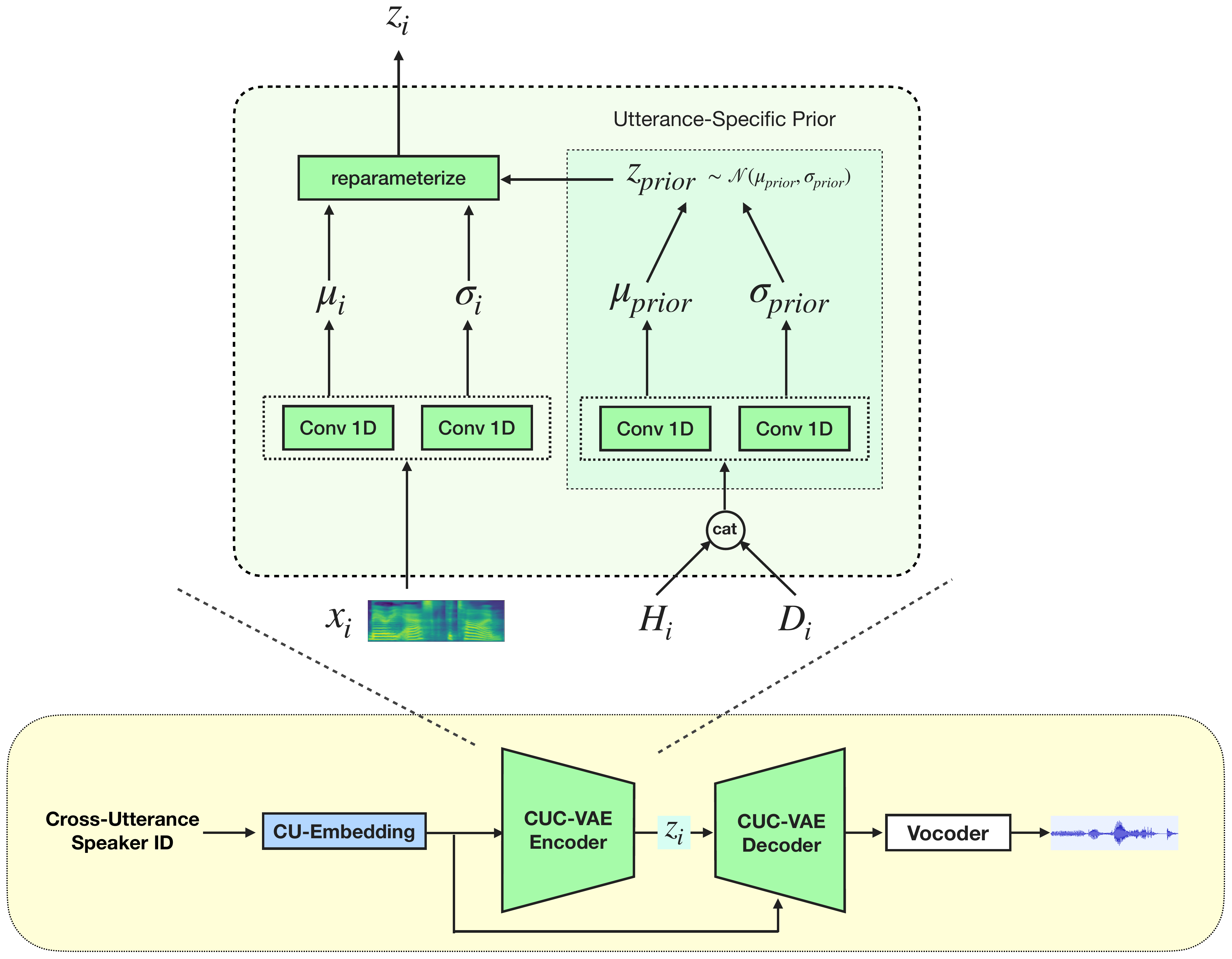}
\caption{A comprehensive overview of the practical CUC-VAE TTS algorithm. 
}
\label{fig:TTS}
\end{figure}

This section presents a detailed exploration of the practical implementation of our proposed algorithm for Text-to-Speech (TTS) tasks, referred to as the CUC-VAE TTS. This algorithm is a specific application of our CUC-VAE S2 framework, designed for TTS tasks. It harnesses the capabilities of the CUC-VAE to model prosody from neighboring utterances, thereby enhancing the naturalness and expressiveness of the synthesized speech.
Fig.~\ref{fig:TTS} illustrates the architecture of the CUC-VAE TTS algorithm, with the preprocessing module intentionally omitted for clarity. The preprocessing module's primary function in the CUC-VAE TTS algorithm is to extract the mel-spectrogram from the waveform in the dataset.

The subsequent discussion will delve into the details of the CUC-VAE TTS algorithm. We propose the CU-enhanced CVAE to overcome the lack of prosody variation and the inconsistency between the standard Gaussian prior distribution sampled by the VAE-based TTS system and the true prior distribution of speech.
Excluding the omitted preprocessing module, the CUC-VAE consists of a CU-embedding module, an encoder module, as illustrated in Fig.~\ref{fig:TTS}, and a decoder module derived from FastSpeech 2. The CU-embedding module's function is to extract potential prosody information from neighborhood utterances, speaker ID, and the current sentence.
The CU-embedding module of the system employs a Transformer to learn the current representation of the utterance, with the dimension of phoneme embeddings and the size of the self-attention set to 256. The "BERT\_BASE" configuration was utilized, consisting of 12 Transformer blocks with 12-head attention layers and a hidden size of 768. The BERT model and associated embeddings were kept fixed during training. Furthermore, information regarding different speakers was incorporated into the Transformer output using a 256-dim embedding layer.

Fig.~\ref{fig:TTS} describes the details of the CUC-VAE encoder. The utterance-specific prior in the encoder aims to learn the prior distribution $\bm z_p$ from the CU-embedding output $\bm H$ and predicts duration $\bm D$. For convenience, the subscript $i$ is omitted in this subsection.
The posterior module in the encoder takes as input reference mel-spectrogram $\bm x$, then models the approximate posterior $\bm z$ conditioned on utterance-specific conditional prior $\bm z_p$. Sampling is done from the estimated prior by the utterance-specific prior module and is reparameterized as:
\begin{equation}
\label{eq:sample_z}
\bm{z} = \bm \mu \oplus \bm \sigma \otimes \bm{z}_{p},
\end{equation}

The element-wise addition and multiplication operations are denoted by $\oplus$ and $\otimes$, respectively. The variable $\bm z_{p}$ is sampled from the utterance-specific prior that has been learned. The re-parameterization can be expressed as follows:
\begin{equation}
\label{eq:sample_zp}
\bm z_{p} = \bm \mu_{p} \oplus \bm \sigma_{p} \otimes \bm \epsilon
\end{equation}

The mean $\bm \mu_{p}$ and variance $\bm \sigma_{p}$ are learned from the utterance-specific prior module. The variable $\bm \epsilon$ is sampled from a standard Gaussian distribution $\mathcal{N}(0,1)$.

By substituting Eq.\ref{eq:sample_zp} into Eq.\ref{eq:sample_z}, the complete sampling process can be described by the following equation:
\begin{equation}
\label{eq:sample_p_new}
\bm z = \bm \mu \oplus \bm \sigma \otimes \bm \mu_{p} \oplus \bm \sigma \otimes \bm \sigma_{p}\otimes \bm \epsilon.
\end{equation}
During the inference phase, sampling is performed from the utterance-specific conditional prior distribution $\mathcal{N}(\bm \mu_{p},\bm \sigma_{p})$ that has been learned from the CU-embedding, instead of using a standard Gaussian distribution $\mathcal{N}(0,1)$. For simplicity, we can formulate the data likelihood calculation as follows, where the intermediate variable utterance-specific prior $\bm z_{p}$ from $\bm D,\bm H$ to obtain $\bm z$ is omitted:
\begin{equation}
\label{eq:our_likehood}
p_\theta(\bm x \mid \bm{H}, \bm{D})=\int p_\theta(\bm x \mid \bm z, \bm{H}, \bm{D})p_\phi(\bm z\mid \bm{H}, \bm{D}) d \bm z,
\end{equation}
In Eq.~\ref{eq:our_likehood}, $\phi$ and $\theta$ represent the parameters of the encoder and decoder modules in the system, respectively.

The decoder employed in the CU-enhanced CVAE is derived from FastSpeech 2. Initially, a projection layer is incorporated to map the latent variable $\bm z$ to a high-dimensional space to facilitate its addition to $\bm H$. Subsequently, the decoder module is used to convert the hidden sequence into a mel-spectrogram sequence through parallelized computation.

As shown in Fig.~\ref{fig:TTS}, four 1D-convolutional (1D-Conv) layers with kernel sizes of 1 were employed to predict the mean and variance of 2-dim latent features. Then, the sampled latent feature was converted to a 256-dim vector by a linear layer. The length regulator in FastSpeech 2's duration predictor, which consisted of two 1D convolutional blocks with ReLU activation, followed by layer normalization and an additional linear layer to predict the length of each phoneme, was adapted to take in the outputs of the CU-embedding module. Each convolutional block was comprised of a 1D-Conv network with ReLU activation, followed by layer normalization and a dropout layer. Four feed-forward Transformer blocks were used by the decoder to transform hidden sequences into an 80-dim mel-spectrogram sequence.

Finally, the vocoder HifiGAN~\cite{hifigan} was fine-tuned for 1200 steps on an open-sourced, pre-trained version of "UNIVERSAL\_V1" to synthesize a waveform from the predicted mel-spectrogram.

\subsection{CUC-VAE SE Algorithm for Speech Editing}
\label{sec:CUC_VAE_SE}

This section is dedicated to the exploration of the practical application of our proposed algorithm for speech editing tasks, known as the CUC-VAE SE. This algorithm is a specialized adaptation of our CUC-VAE S2 framework, specifically tailored to address speech editing tasks.
In this study, our primary focus is on the partial editing setting, where the requirement is to edit only a portion of the waveform rather than the entire waveform. 
Consequently, the CUC-VAE SE is designed to leverage the strengths of the CUC-VAE to model prosody from both adjacent utterances and the neighboring waveform of the editing part. This approach significantly enhances the quality and expressiveness of the edited speech, ensuring a seamless integration of the edited part with the rest of the waveform.

\begin{figure*}
\centering
\includegraphics[width=0.9\linewidth]{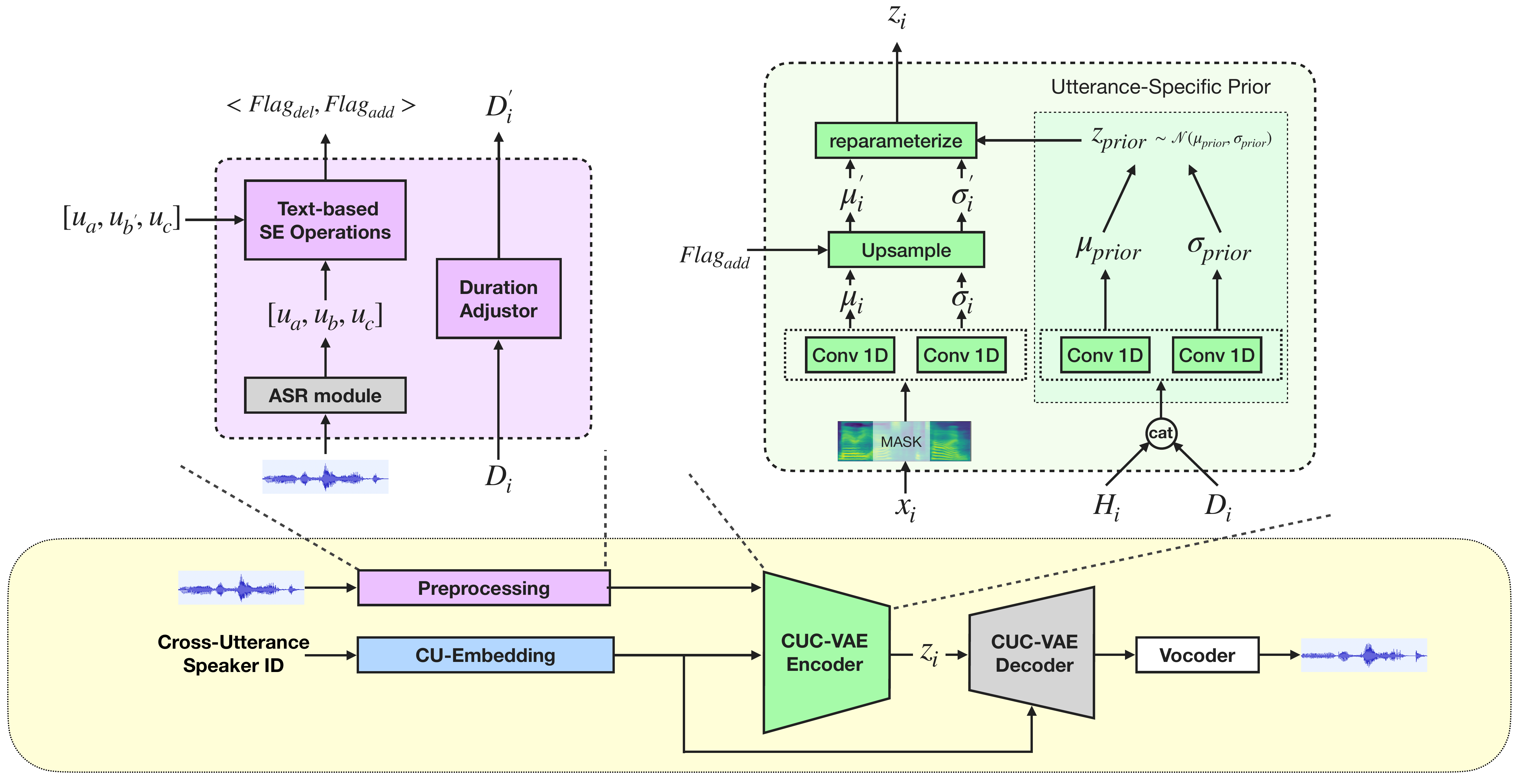}
\caption{A comprehensive overview of the practical CUC-VAE SE algorithm.}
\label{fig:SE}
\end{figure*}

In this study, we concentrate on three primary speech editing operations: deletion, insertion, and replacement. Assuming the original utterance transcript of the speech as $[\bm{u}_a,\bm{u}_b,\bm{u}_c]$, the modified utterance can be represented as $[\bm{u}_a,\bm{u}_{b^{\prime}},\bm{u}_c]$, where $\bm{u}_{b^{\prime}}$ denotes the modified segment while $\bm{u}_{a}$ and $\bm{u}_c$ remain unchanged. The corresponding phonemes translated by G2P can be denoted as $\bm{p}_i = [\bm{p}_a,\bm{p}_b,\bm{p}_c]$, and the original speech's mel-spectrogram can be denoted as $\bm{x}_i=[\bm{x}_a , \bm{x}_b , \bm{x}_c]$. Here, $\bm{x}_i$ contains a sequence of frame-level mel-spectrogram for $i \in \{a, b, c\}$.

Although there are three primary speech editing operations: deletion, insertion, and replacement, the replacement operation can be considered as a deletion followed by an insertion.
Therefore, we can use two flags, namely $Flag_{del}$ and $Flag_{add}$, to indicate the location of deletion and addition, as shown in Fig.~\ref{fig:SE}.

\paragraph{Deletion}
The deletion operation allows the user to remove a segment of the speech waveform that corresponds to a particular set of words. After the deletion, the target utterance to be synthesized becomes $[\bm{u}_a,\bm{u}_c]$, with $\bm{u}_b$ representing the segment to be removed. The comparison between the original and edited utterances provides the deletion indicator, denoted by $Flag_{del}$, which is used to guide the editing of the mel-spectrogram. Specifically, $Flag_{del}$ is defined as follows
$Flag_{del} = [\bm{0}_a,\bm{1}_b,\bm{0}_c],$ where $\bm_{0}$ and $\bm_{1}$ denote zero and one vectors, respectively.

\paragraph{Insertion and Replacement}
Unlike the deletion operation, the target synthesized speech after insertion or replacement is based on the edited utterance $[\bm{u}_a, \bm{u}_{b^{\prime}}, \bm{u}_c]$, where $\bm{u}_{b^{\prime}}$ denotes the content to replace $\bm{u}_b$. It is worth noting that the insertion process can be treated as a special case where $\bm{u}_b = \bm{p}_b = \bm{x}_b = \varnothing$. Correspondingly, we can define the addition indicator as 
$Flag_{add} = [\bm{0}_a,\bm{1}_{b^{\prime}},\bm{0}_c].$

As shown in Fig.~\ref{fig:SE}, the CUC-VAE synthesizer is utilized to generate the mel-spectrogram $\bm{x}_{b^{\prime}}$, based on the reference mel-spectrogram $[\bm{x}_a , \bm{x}_c]$ and the neighborhood utterances. 
In this process, two one-dimensional convolutions are employed to learn the mean $\bm{\mu}$ and variance $\bm{\sigma}$. Referring to $Flag_{add}$, the corresponding position of $\bm{\mu}$ and $\bm{\sigma}$ are updated by adding 0s and 1s, resulting in $\hat{\bm \mu} = [\bm{\mu}_a, \bm{0}_{b^{\prime}},\bm{\mu}_c]$ and $\hat{\bm \sigma} = [\bm{\sigma}_a , \bm{1}_{b^{\prime}},\bm{\sigma}_c]$, respectively. This allows the generation of speech for the edited region from the utterance-specific prior distribution, while the unmodified regions are sampled from the actual audio and the utterance-specific prior distribution. During the training phase, the real audio that has been edited is not available. Therefore, certain segments of audio are masked, and the same content is restored to simulate the editing scenario. As a result, $b^{\prime}$ is set to $b$.

To enhance the coherence and contextual relevance of the output generated by the modified CUC-VAE module, we have embedded two specific additional modules into the CUC-VAE S2 framework, i.e., Upsampling in CUC-VAE Encoder and Duration Adjustor in preprocessing module. To achieve a smoother editing boundary, the values of $\hat{\bm \mu}$ and $\hat{\bm \sigma}$ are convolved using one-dimensional convolution to obtain $\bm \mu^{\prime}$ and $\bm \sigma^{\prime}$. By employing this approach, the module can perform sampling from the estimated prior distribution and can be further re-parameterized as follows:
$
\label{eq:sample_z_se}
\bm{z} = \bm \mu^{\prime} \oplus \bm \sigma^{\prime} \otimes \bm{z}_{p},
$

The re-parameterization formula and ELBO objective in this module are similar to the original CUC-VAE module. 
Specifically, as shown in Fig.~\ref{fig:SE}, the upsampling is achieved by an additional upsampling layer to ensure that the predicted sequence length matched the phoneme sequence length after editing and to enhance the naturalness of the synthesized audio. 

In addition, to efficiently leverage the duration information obtained from the original audio, a similar approach to that used in \cite{EditSpeech, a3t} is adopted. Consequently, the phoneme duration within the edited region is adjusted by multiplying it with the ratio of the duration of the original audio to the predicted duration of the unedited area in the audio to obtain $\bm{D}^{\prime}_i$. Following the duration predictor and adjustor, the predicted duration is rounded to the nearest integer value.

The main difference between these two modules is that while carrying out the inference stage, the sampling process remains coherent with the training stage, sampling from the masked mel-spectrogram conditioned on the utterance-specific conditional prior.
Besides, to accurately replicate the actual editing scenario, we selected the part to be masked by taking a word instead of a phoneme as a unit. Additionally, to achieve a balance between the system's ability to learn and predict audio information, we set the shielding rate to 50\%, a value that has been found to be effective in previous studies\cite{a3t}.

On the other hand, the reconstruction of a waveform from a given transcript for each masked phoneme requires the use of a loss function for the acoustic model. The most commonly used loss function is the mean absolute error (MAE) between the reconstructed and original mel-spectrogram, with the loss being computed only on the masked segments, similar to the BERT model.

During the training process, the input reference mel-spectrogram only includes the unmasked part. To give more emphasis on the masked part, it is reasonable to increase the loss weight of this area. However, during the inference process, the goal is to synthesize naturally coherent audio with rhythms that conform to \changes{the modified} text context. Therefore, setting the loss weight of the unmasked area to zero is not appropriate in the case of speech editing.

To balance the two objectives of approaching the original audio and the context of the newly modified transcript, we propose setting the loss ratio of the masked and unmasked parts to $\lambda=1.5$ in the experiment. In this way, we expect to achieve a satisfactory outcome that is consistent with both objectives.
\begin{equation}
\begin{aligned}
\mathcal{L}{mel} &= \frac{1}{\# \text { of frames}} (\sum{i \in \text{unmask}} | x_i^{pred}-x_i^{target}| \\
&+ \lambda \sum_{i \in \text{mask}} | x_i^{pred}-x_i^{target}|)
\end{aligned}
\label{obj:se}
\end{equation}
The findings of the subsequent experiments indicate that increasing the weight of the loss function associated with the reconstructed mel-spectrogram of the masked part, in comparison to other weight settings, can lead to the generation of a more natural and coherent synthesized sound.

The remaining modules, including the CU-embedding, CUC-VAE Decoder, and Vocoder, are implemented in the same manner as in the CUC-VAE TTS algorithm.

\section{Experiments}
\label{sec:exp_results}
In this section, we present a series of experiments designed to assess the efficacy of our proposed speech editing system. We begin by outlining the experimental setup, including the dataset and evaluation metrics used. Subsequently, we compare the naturalness and prosody diversity of speech synthesized using our CUC-VAE model with that of FastSpeech 2 and other VAE techniques. We then evaluate the naturalness and similarity of the audio generated by our system against that of EditSpeech~\cite{EditSpeech}, using both partial and entire inference. An ablation study is also conducted to examine the impact of limiting contextual information on our system's performance, as measured by both the mean opinion score (MOS) and reconstruction performance. Furthermore, we investigate the effect of biased training on reconstruction performance. Finally, we present two case studies that illustrate the variations in prosody with different cross-utterance information and the influence of entire inference and biased training, respectively. Audio samples from these experiments can be accessed on our demo page (\url{https://bit.ly/CUC-VAE-SG}).

\begin{table}[t]
    \centering
    \caption{Sample naturalness and diversity results on LibriTTS dataset.
    FS2 is an abbreviation for the baseline model, FastSpeech 2.
    }
    \resizebox{\linewidth}{!}{
    \begin{tabular}{ccccccc}
    \toprule[1pt]
    \multirow{2}{*}{Method}&
    \multirow{2}{*}{{MOS}}&
    \multirow{2}{*}{{FFE}}&
    \multirow{2}{*}{{MCD}}&
    \multirow{2}{*}{{WER}}&
    \multicolumn{2}{c}{{Prosody Std.}}\\
 \cmidrule(lr){6-7}
 &&& && $\bm{F}_0$ & $\bm{E}$  \\   
    \hline
        GT & 4.10 $\pm$ 0.07 & - & - & 3.1 & - & -\\
        GT (Mel+HifiGAN) & 4.03 $\pm$ 0.07 & 0.17 & 4.65 & 3.9 & - & -\\
        \hdashline
         FS2 &3.53 $\pm$ 0.08 & 0.58 & 6.32 &6.0&$2.13\times 10^{-13}$ &$7.22\times 10^{-7}$\\
         FS2 +Global VAE &3.59 $\pm$ 0.08 & 0.45 & 6.27 &10.8&2.01 & 0.0054 \\
         FS2+Fine-grained VAE &3.43 $\pm$ 0.08 & 0.35 & 6.28 & 5.6 & 63.64 & 0.0901\\ 
         CUC-VAE-TTS &\textbf{3.63} $\bm\pm$ \textbf{0.08} &\textbf{0.34} &\textbf{6.04} &\textbf{5.5} & 30.28& 0.0217\\
    \bottomrule[1pt]
    \end{tabular}
    }
    \label{tab:nd}
\end{table}

\subsection{Environmental Setting}
In this study, we performed experiments using the LibriTTS dataset~\cite{libritts} , which is a multi-speaker dataset comprising the train-clean-100 and train-clean-360 subsets. These subsets contain 245 hours of English audiobooks recorded by 1151 speakers, consisting of 553 female and 598 male speakers. \changes{The original LibriTTS dataset, structured by audiobooks, inherently includes adjacent sentences, facilitating the extraction of contextual information.
Specifically, to enrich the dataset with contextual information, we reprocess the LibriTTS dataset by integrating surrounding $l$ utterances using a sliding window technique: $<(speech, text)_{-l}, ..., (speech, text)_{0}, ..., (speech, text)_{l}>$.}
We randomly selected 90\%, 5\%, and 5\% of the data from the dataset for the training, validation, and testing sets, respectively. All audio clips were resampled at a rate of 22.04 kHz.

To evaluate the performance of the proposed method, both subjective and objective tests were conducted. The subjective test involved 20 volunteers who were asked to assess the naturalness and similarity of 15 synthesized speech samples using a 5-point mean opinion score (MOS) evaluation. The MOS results were analyzed using 95\% confidence intervals and p-values.

\begin{table}[t]
    \centering
    \caption{\changes{Sample naturalness and diversity comparison between Grad-TTS and our proposed CUC-VAE-TTS on LibriTTS dataset.
    Three metrics are reported, namely FFE, MCD, WER.}}
    \begin{tabular}{cccc}
        
    \toprule[1pt]
        {Model} & {FFE} & {MCD} & {WER} \\
        \midrule[1pt]
        {Grad-TTS} & \textbf{0.2734} & 8.3102 & 13.479 \\
        {CUC-VAE-TTS} & 0.2769 & \textbf{6.9661}& \textbf{6.2866} \\
        
    \bottomrule[1pt]
    \end{tabular}
\label{tab:nd1}
\end{table}

For the objective evaluation, two metrics were used: F0 frame error (FFE)~\cite{ffe} and Mel-cepstral distortion (MCD)~\cite{mcd}. FFE was utilized to evaluate the accuracy of the F0 track reconstruction, which combined the Gross Pitch Error (GPE) and the Voicing Decision Error (VDE). More specifically, GPE measures the difference between the predicted and reference F0 values, while VDE measures the accuracy of voicing decision (i.e., whether a frame is voiced or unvoiced). On the other hand, MCD measures the spectral distortion between the predicted and reference mel-spectrograms. These objective metrics were used to evaluate the performance of different VAEs and different settings of loss weights.

In detail,
$$
\begin{aligned}
\operatorname{FFE} &=\frac{\# \text { of error frames }}{\# \text { of total frames }} \times 100 \% \\
&=\frac{N_{U \rightarrow V}+N_{V \rightarrow U}+N_{\operatorname{F0E}}}{N} \times 100 \% .
\end{aligned}
$$
where $N_{U \rightarrow V}$ and $N_{V \rightarrow U}$ are the numbers of unvoiced/voiced frames classified as voiced/unvoiced frames, $N$ is the number of the frames in the utterance, and $N_{\operatorname{F0E}}$ is number of frames for which
$$
\left|\frac{F 0_{i, \text { estimated }}}{F 0_{i, \text { reference }}}-1\right|> 20 \%
$$
where $i$ is the frame number.

Besides, MCD evaluated the timbral distortion, computing from the first 13 MFCCs in our trials.
$$
\operatorname{MCD}(\bm{y}, \hat{\bm{y}})=\frac{10 \sqrt{2}}{\ln 10}\|\bm{y}-\hat{\bm{y}}\|_2 \quad \text{(dB)}
$$
where $\bm{y}$ and $\hat{\bm{y}}$ are the MFCCs of original and reconstructed waveform. A coefficient was utilized to convert the Mel-cepstral distortion (MCD) units into decibels. The MCD represents the difference between the synthesized and natural mel-spectrogram sequences, and a smaller MCD value indicates a closer resemblance to natural speech, thus reflecting naturalness.

In addition to assessing naturalness, we also reported word error rates (WER) from an automatic speech recognition model, which provides a measure of the intelligibility and consistency between synthetic and real speech. To this end, an attention-based encoder-decoder model trained on Librispeech 960-hour data was utilized. Notably, the model is open-sourced and can be accessed at \url{http://bitly.ws/uMKv}.

\begin{table*}[t]
    \caption{The significance analysis of our system employing entire inference as compared to "Mel$\_$cut" and "Wave$\_$cut" in terms of the MOS scores for naturalness and similarity.}
    \label{tab:pvalue}
    
    \centering
    \begin{tabular}{ccccccccc}
    \toprule[1pt]
    \multirow{2}{*}{Method} &
    \multicolumn{2}{c}{Insert} & \multicolumn{2}{c}{Replace} & \multicolumn{2}{c}{Delete} & \multicolumn{2}{c}{Reconstruct} \\
    \cmidrule(lr){2-3} \cmidrule(lr){4-5} \cmidrule(lr){6-7} \cmidrule(lr){8-9}
     & Nat. & Sim. & Nat. & Sim. & Nat. & Sim. & Nat. & Sim. \\
    \hline
         CUC-VAE-SE vs. Mel$\_$cut &
                    0.0662  & 0.793 & 
                    0.0294  & 0.771 &
                    0.0168  & 0.298 &
                    0.0525  & 0.691 \\
         CUC-VAE-SE vs. Wave$\_$cut   & 
                    0.0219   & 0.135 &
                    0.0163   & 0.287 &
                    0.369   & 0.310 &
                    0.0564  & 0.143 \\
    \bottomrule[1pt]
    \end{tabular}
\end{table*}

\begin{table}[ht]
    \caption{The results of subjective evaluations of naturalness and similarity for the EditSpeech and our system approaches using both partial and entire inference methods.}
    \label{tab:partial/entire_inference}
    
    \centering
    \begin{tabular}{ccccccc}
    \toprule[1pt]
    \multirow{2}{*}{Method} &
    \multicolumn{2}{c}{Insert} & \multicolumn{2}{c}{Replace} & \multicolumn{2}{c}{Delete} \\
    \cmidrule(lr){2-3} \cmidrule(lr){4-5} \cmidrule(lr){6-7} 
     & Nat. & Sim. & Nat. & Sim. & Nat. & Sim. \\
    \hline
         Wave$\_$cut& 2.93      & 3.76 &
                      2.82      & 3.50 &
                      3.25      & 3.82 \\
         \hdashline
         EditSpeech (Mel$\_$cut) & 2.35      & 3.21      &
                      2.47      & 3.36      &
                      2.82      & \bf{3.81}     \\
         CUC-VAE-SE (Mel$\_$cut) & 3.11      & \bf{3.57} &
                      2.97      & 3.41      &
                      2.82      & \bf{3.81}     \\
         CUC-VAE-SE & \bf{3.37} & 3.56      & 
                      \bf{3.39} & \bf{3.43} &
                      \bf{3.37} & 3.67      \\
    \bottomrule[1pt]
    \end{tabular}

\end{table}
\subsection{Sample Naturalness and Diversity}

We measure the sample naturalness and intelligibility using MOS and WER. 
Complementary to the naturalness, the diversity of generated speech from the conditional prior was evaluated by comparing the standard deviation of $E$ and $F_0$ similar to \cite{sun2020finegrainedvae}.

The results were shown in Table.~\ref{tab:nd}. 
Compared to the global VAE and fine-grained VAE, the proposed CUC-VAE received the highest MOS and achieved the lowest FFE, MCD, and WER.
Although both $F_0$ and $E$ of the CUC-VAE TTS module were lower than the baseline + fine-grained VAE, the proposed system achieved a clearly higher prosody diversity than the baseline and baseline + global VAE systems.
\changes{
Additionally, Table~\ref{tab:nd1} presents a comparative analysis of sample naturalness and diversity between Grad-TTS~\cite{popov2021grad} and our proposed CUC-VAE-TTS method using the LibriTTS dataset. Notably, our approach demonstrates a significant decrease in both MCD and WER, indicating improved prosody modeling and synthesis quality. It is important to note that the variations in the listed metrics between Table~\ref{tab:nd} and Table~\ref{tab:nd1} arise from differences in the evaluation sets between our method and the official GradTTS. Specifically, we use the same 122 test samples for consistency. These results underscore the efficacy of our proposed method in capturing diverse and context-sensitive prosody within the TTS domain.
}
The fine-grained VAE achieved the highest prosody variation as its latent prosody features were sampled from a standard Gaussian distribution, which lacks the constraint of language information from both the current and the neighbouring utterances. This caused extreme prosody variations to occur which impaired both the naturalness and the intelligibility of synthesized audios.
As a result, the CUC-VAE TTS module was able to achieve high prosody diversity without hurting the naturalness of the generated speech. In fact, the adequate increase in prosody diversity improved the expressiveness of the synthesized audio, and hence increased the naturalness.
\begin{table}[h]
    \caption{The subjective and objective results of the reconstruction performance of EditSpeech and our system using partial and entire inference.}
    \label{tab:partial/entire_inferenc_reconstruction}
    
    \centering
    \begin{tabular}{cccccc}
    \toprule[1pt]
    Method & Nat. & Sim. & {FFE} & {MCD} & {WER}\\
    \hline
         GT & 4.56 & - & - & - & 3.124\\
         GT (Mel+HifiGAN) & 4.39 & 4.68 & 0.170 & 4.651 & 3.887\\
         \hdashline
         EditSpeech & 3.14       & 3.80  & 0.372 & 6.345 & 6.702\\
         CUC-VAE-SE (Mel$\_$cut) & 3.66  & \bf{3.91} & \bf{0.326} & \bf{5.957} & \bf{5.174}\\
         CUC-VAE-SE & \bf{3.90}  & 3.83  & 0.327 & 6.657 & 5.377\\
    \bottomrule[1pt]
    \end{tabular}
\end{table}
\subsection{Partial vs. Entire Inference}
To compare the performance of partial inference versus entire inference in the context of speech editing, several experiments were conducted on different systems: 
1) {GT}, the ground truth audio; 
2) {GT (Mel+HifiGAN)}, the ground truth audio converted to mel-spectrogram and then back to audio using HifiGAN vocoder;
3) {Wave$\_$cut}, a modified version of the waveform obtained by manually cutting and reinserting a specific region;
4) {EditSpeech}~\cite{EditSpeech}, an approach based on partial inference and bidirectional fusion to improve prosody near boundaries;
5) {CUC-VAE-SE (Mel$\_$cut)}, an approach that involves cutting the modified region from the generated mel-spectrogram and inserting it back to the original mel-spectrogram using a forced aligner; 
6) {CUC-VAE-SE}, an approach that regenerates a complete mel-spectrogram from the entire sentence to be edited and then uses HifiGAN vocoder to generate the complete waveform; 

Note that the ground truth audio samples used in this study do not contain any edited audio. Therefore, the systems designated as {GT} and {GT (Mel+HifiGAN)} were used to evaluate the reconstruction performance of the audio signals. To assess the effectiveness of the editing operations, we manually spliced the audio waveform and used the MOS similarity score of the resulting system, designated as {Wave$\_$cut}, as an indicator of the upper bound performance.

Table~\ref{tab:partial/entire_inference} presents the MOS scores for naturalness and similarity obtained from the subjective evaluations of the various editing operations. Note that the EditSpeech approach is solely capable of partial inference, whereby segments of the real mel-spectrogram are combined to generate the modified speech. Consequently, the results of the deletion operation are uniform across different editing systems when partial inference is utilized.

\begin{table}[t]
    \caption{Objective metrics and subjective results of the performance of our system with various VAEs.}
    \label{tab:ablation_study1_2}
    
    \centering
    \resizebox{\linewidth}{!}{
    \begin{tabular}{ccccccccccc}
    \toprule[1pt]
    \multirow{2}{*}{Method} & \multicolumn{3}{c}{Objective Metrics} & \multicolumn{4}{c}{Subjective Results} \\
    \cmidrule(lr){2-4} \cmidrule(lr){5-8}
    & {FFE} & {MCD} & {WER} & {Insert} & {Replace} & {Delete} & {Reconstruct} \\
    \hline
         GT & - & - & 3.124 & - & - & - & 4.56$\pm$0.06 \\
         GT (Mel+HifiGAN) & 0.170 & 4.651 & 3.887 & - & - & - & 4.39$\pm$0.05 \\
         \hdashline
         Baseline1 & 0.371 & 6.919 & 7.404 & 2.91$\pm$0.11 & 2.87$\pm$0.10 & 3.31$\pm$0.15 & 2.90$\pm$0.08 \\
         Baseline2 & 0.333 & 6.750 & 5.503 & 3.43$\pm$0.12 & 3.11$\pm$0.09 & 3.52$\pm$0.14 & 3.45$\pm$0.10 \\
         Baseline3 & 0.332 & 6.697 & 5.392 & 3.79$\pm$0.11 & 3.37$\pm$0.13 & 3.92$\pm$0.14 & 3.62$\pm$0.09 \\
         CUC-VAE-TTS/SE & \bf{0.327} & \bf{6.657} & \bf{5.377} & \bf{3.93$\pm$0.10} & \bf{3.43$\pm$0.13} & \bf{4.29$\pm$0.12} & \bf{3.90$\pm$0.10} \\
    \bottomrule[1pt]
    \end{tabular}}
\end{table}

As shown, our model employing entire inference achieved the highest score for naturalness across all editing operations. Notably, a significant gap was observed in the replacement operation, where the speech editing models based on partial reasoning faced challenges in dealing with intonation conversion. On the other hand, the MOS score for naturalness of "Mel$\_$cut" in the deletion operation was relatively low since its performance heavily relies on the accuracy of the forced aligner, particularly when short words are deleted. In such cases, its performance may be inferior to manually conducted waveform-based deletions.

Regarding the MOS scores for similarity, our system based on entire inference demonstrated comparable performance to "Mel$\_$cut" based on partial inference, and outperformed EditSpeech in both insertion and replacement operations. Additionally, it exhibited a similarity score close to that of "Wave$\_$cut", which served as the upper bound indicator for similarity, with the maximum difference being approximately 0.2.

Table~\ref{tab:pvalue} reports the p-values obtained from the statistical analysis of the MOS scores for naturalness and similarity. The results indicate that our model using entire inference outperforms both "Mel$\_$cut" and "Wave$\_$cut" in terms of naturalness, as the p-values are significant. Meanwhile, there was no significant difference in similarity between entire inference and the two partial inference methods.
The only exception was in the case of the deletion operation, where there was no significant difference in naturalness between our model using entire inference and ``Wave$\_$cut".

Table~\ref{tab:partial/entire_inferenc_reconstruction} provides evidence of the efficacy of our mask CU-enhanced CVAE module in reconstructing the mel-spectrogram. 
Note that lower objective results indicate higher similarity to the original audio. 
As expected, CUC-VAE-SE(Mel\_cut), which involves copying the 50\% unedited area of the real mel-spectrogram, showed superior reconstruction performance in terms of MCD. 
Despite not directly copying the unedited area's real mel-spectrogram, our entire inference-based system demonstrated comparable similarity performance to EditSpeech.
Also, CUC-VAE-SE system outperformed EditSpeech in naturalness, FFE, and WER with a clear margin. 

As a result, our CUC-VAE-SE system has demonstrated significantly better subjective naturalness and similarity scores as well as multiple objective metrics in speech reconstruction and multiple speech editing tasks than EditSpeech. Furthermore, it has been demonstrated that utilizing entire inference yields significantly better naturalness compared to partial inference, while the similarity of the inferred results to the true mel-spectrogram or waveform is not significantly different. This showcases the reliable reconstruction capability and rich expressiveness of our proposed system.

\begin{table}[ht]
    \caption{Reconstruction performance of our system trained using different loss ratios. Loss ratio is unmasked: masked part.}
    \label{tab:ablation_study3}
    
    \centering
    \begin{tabular}{ccccc}
    \toprule[1pt]
    Method & Loss ratio &
    {FFE} & {MCD} & {WER}\\
    \hline
         \multirow{5}{*}{CUC-VAE-SE} & 1:1 & 0.437 & 6.931 & 5.406\\
          & 1:1.5 & 0.327 & \bf{6.657} & \bf{5.377}\\
          & 1:2 & \bf{0.326} & 6.697 & \bf{5.377}\\
          & 1:5 & 0.434 & 6.824 & 5.525\\
          & 0:1 & 0.457 & 7.596 & 14.808\\
    \bottomrule[1pt]
    \end{tabular}
\end{table}

\subsection{Ablation Study on Different VAEs}
In this section, we investigate the impact of using different VAEs on the performance of our system.
We conduct a comparative analysis of the reconstruction performance and MOS scores of the synthesized audio across various systems, including those used in the previous experimental settings. We further compare the performance of our system with that of several baseline models, namely
1) {Baseline1}, uses a fine-grained VAE instead of CUC-VAE; 
2) {Baseline2}, uses a CVAE without the context embeddings (i.e., $l=0$); 
3) {Baseline3}, uses CUC-VAE with 2 neighbouring utterances (i.e., $l=2$); 
4) {Our system}, use CUC-VAE with 5 neighbouring utterances (i.e., $l=5$).
\changes{In this context, $l$ represents the number of neighboring utterances on each side of the current utterance, encompassing both $l$ preceding and $l$ subsequent utterances of current one.}

    

    

As presented in Table~\ref{tab:ablation_study1_2}, the introduction of semantic restriction of edited text and context embeddings in succession resulted in a steady increase in both the MOS for naturalness of edited and reconstructed waveforms and objective scores for reconstruction performance. 
The score metrics also indicate that the inclusion of more cross-utterances could enhance the system's reconstruction capability. However, the improvements observed by using more than five neighboring utterances were negligible. Consequently, the subsequent experiments were carried out using $L=5$. 
These results suggest that the CU-embedding and mask CUC-VAE module played a pivotal role in producing more coherent audio.

\begin{table}[h]

    \caption{Performance comparison of CUC-VAE-TTS with different embedding foundation models: Word2Vec, Glove and Bert.}
    \label{tab:semantic_embeddings}
    
    \centering
    \begin{tabular}{cccc}
        \toprule
        {Method} & {FFE} & {MCD} & {WER} \\
        \midrule
        {w/ Word2Vec} & 0.3271 & 7.1630 & 5.4404 \\
        {w/ GloVe} & 0.2814 & 7.3122 & 5.5694 \\
        {w/ Bert} & 0.3400 & 6.0400 & 5.5000 \\
        \bottomrule
    \end{tabular}

\end{table}
\subsection{Ablation Study on the Degree of Biased Training}

In this section, another ablation study based  on reconstruction performance was conducted to investigate the impact of attention intensity on the masked mel-spectrogram. 
Specifically, the study evaluated a system trained with different ratios of attention to the masked and unmasked areas with ratio=1:1, 1:1.5, 1:2, 1:5, and 0:1. The ratio=1:1 represents the use of a normal loss function that treats the masked and unmasked areas equally, while ratio=0:1 represents the application of partial inference used by existing speech editing systems that only focuses on the masked areas.

The purpose of this experiment is to identify a parameter that enables the generated edited audio to be coherent and retain the prosody characteristics of the original audio while using either the common TTS overall reasoning or the partial reasoning of existing speech editing systems.

\begin{figure*}[ht!]
\centering
\subfigure[Target]{\label{fig:subfig:cs2:target}
\includegraphics[scale=0.31]{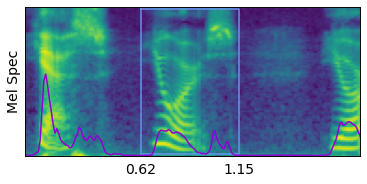}}
\subfigure[EditSpeech]{\label{fig:subfig:cs2:editspeech}
\includegraphics[scale=0.31]{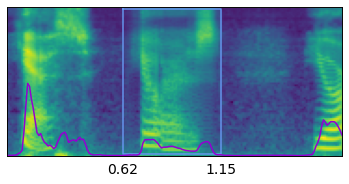}}
\subfigure[Our system(Unbiased)]{\label{fig:subfig:cs2:loss1}
\includegraphics[scale=0.31]{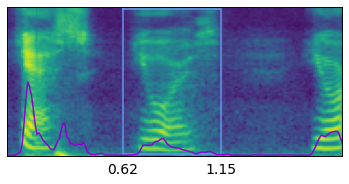}}
\subfigure[Our system(Biased)]{\label{fig:subfig:cs2:oursystem}
\includegraphics[scale=0.31]{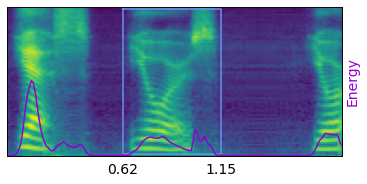}}
\caption{The mel-spectrograms of target speech and speech edited by EditSpeech, our system with unbiased training(loss ratio=1:1), and our system with biased training(loss ratio=1:1.5). The region marked with time (0.62s $\sim$ 1.15s) is the edited region.}
\label{fig:cs2}
\end{figure*}

The results presented in Table~\ref{tab:ablation_study3} indicate that the system trained with a ratio of 1:1.5 attained the lowest values of MCD and WER, with a slightly higher FFE score than the system trained with a ratio of 1:2. These findings suggest that appropriately increasing the attention devoted to the masked mel-spectrogram can effectively enhance the overall quality of the resulting audio output.

\subsection{\changes{Ablation Study on Embedding Models}}
In the following subsection, we delve deeper into examining a variety of pretrained semantic models, including BERT~\cite{bert}, Word2Vec~\cite{mikolov2013efficient}, and GloVe~\cite{glove}. 
Our motive here is to carry out an in-depth evaluation of the impact these models have on the effectiveness of our proposed CUC-VAE-TTS system.
The efficiency of our suggested CUC-VAE-TTS framework, when combined with different base models for embedding like Word2Vec, GloVe, and BERT, is delineated in Table \ref{tab:semantic_embeddings}. 
\lastchanges{The performance of the CUC-VAE-TTS system with different embedding models was evaluated using three key metrics: FFE, MCD, and WER. Our proposed CUC-VAE-TTS with BERT embeddings achieved a significantly lower MCD of 6.0400, indicating superior spectral quality compared to the second-best result of 7.1630 with Word2Vec. However, the FFE and WER values for BERT are slightly higher than those achieved with the other two embedding methods. The ablation studies using Word2Vec and GloVe suggest that while BERT effectively reduces spectral distortion, it may not provide the best performance in F0 track reconstruction accuracy and word accuracy. These findings underscore the trade-offs between embedding models.
}

\subsection{Case Studies}
\subsubsection{The influence of the utterance-specific prior}

To better illustrate how the utterance-specific prior influenced the naturalness of the synthesized speech under a given context, a case study was performed by synthesizing an example utterance, “Mary asked the time”, with two different neighbouring utterances: “Who asked the time? Mary asked the time.” and “Mary asked the time, and was told it was only five.” Based on linguistic knowledge, to answer the question in the first setting, an emphasis should be put on the word “Mary”, while in the second setting, the focus of the sentence is “asked the time”. 

\begin{figure}[h!]
\centering
\subfigure[Who asked the time? \textbf{Mary asked the time.}]{\label{fig:subfig:a}
\includegraphics[width=0.9\linewidth]{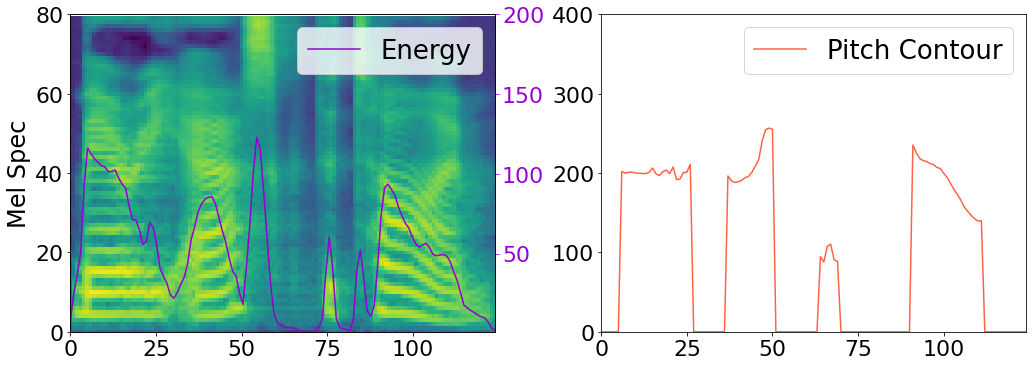}}
\vfill
\subfigure[\textbf{Mary asked the time,} and was told it was only five.]{\label{fig:subfig:b}
\includegraphics[width=0.9\linewidth]{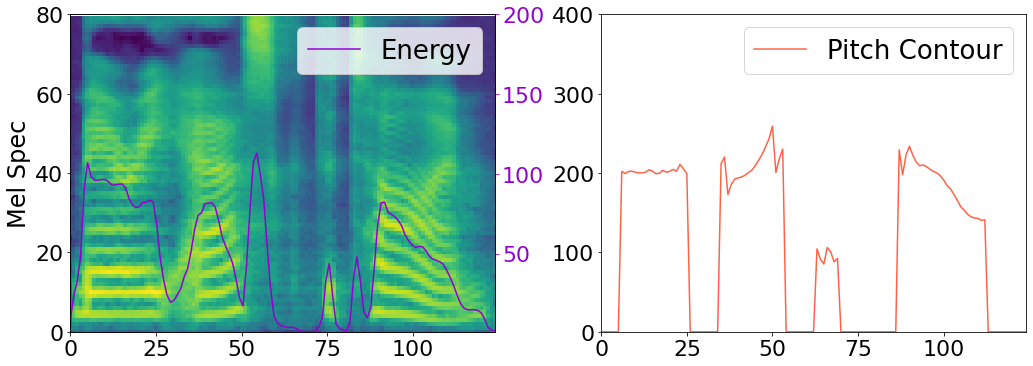}}
\caption{Comparisons between the energy and pitch contour of same text “Mary asked the time" but different neighbouring utterances, generated by CUC-VAE TTS module.}
\label{fig:cs}
\end{figure}

Fig.~\ref{fig:cs} showed the energy and pitch of the two utterances. The energy of the first word “Mary” in Fig.~\ref{fig:subfig:a} changed significantly (energy of “Ma-” was much higher than “-ry”), which reflected an emphasis on the word “Mary”, whereas in Fig.~\ref{fig:subfig:b}, the energy of “Mary” had no obvious change, i.e., the word was not emphasized. 

On the other hand, the fundamental frequency of the words “asked” and “time” stayed at a high level for a longer time in the second audio than the first one, reflecting another type of emphasis on those words which was also coherent with the given context. Therefore, the difference in energy and pitch between the two utterances demonstrated that the speech synthesized by our model is sufficiently contextualized.

\subsubsection{The influence of entire inference and biased training}

To demonstrate the effectiveness of entire Inference and biased training in avoiding unnatural transitions and reconstructing edited and non-edited regions accurately,another case
study was performed by synthesizing an example utterance,
"Yes, yours, your own property".
The reconstruction results of different systems as shown in Fig.~\ref{fig:cs2}, where the area from 0.62s to 1.15s is the edited region that needs to be reconstructed. The text of the masked region is "yours".

Fig.~\ref{fig:subfig:cs2:target} presents the mel-spectrogram and energy contour of the target audio, Fig.~\ref{fig:subfig:cs2:editspeech} is the result of EditSpeech which uses partial inference, and the results of our entire inference approach using unbiased and biased training in Fig.~\ref{fig:subfig:cs2:loss1} and~\ref{fig:subfig:cs2:oursystem}, respectively. 

It can be observed that EditSpeech exhibits distinct segmentation lines at the boundary between the edited and non-edited regions in the mel-spectrogram, whereas the entire inference approach results in more natural transitions and produces reconstructed edited regions that are closer to the target audio. 
Additionally, the energy contour of the edited region in the biased training system is closer to the target audio compared to that of the unbiased training system.

\section{Conclusion}
\label{sec: conclusion}
In this work, we introduce the Cross-Utterance Conditional Variational Autoencoder Speech Synthesis framework, designed to enhance the expressiveness and naturalness of synthesized speech. 
This framework leverages the powerful representational capabilities of pre-trained language models and the re-expression abilities of VAEs. The core component of the CUC-VAE S2 framework is the cross-utterance CVAE, which extracts acoustic, speaker, and textual features from surrounding sentences to generate context-sensitive prosodic features, more accurately emulating human prosody generation.
We further propose two practical algorithms, CUC-VAE TTS for text-to-speech and CUC-VAE SE for speech editing, to address real-world speech synthesis challenges. 
The efficacy of these proposed systems was thoroughly evaluated through a series of comprehensive experiments conducted on the LibriTTS English audiobook dataset. The results of these experiments demonstrated a significant improvement in the prosody diversity, naturalness, and intelligibility of the synthesized speech, thereby validating the effectiveness of our proposed systems.

\bibliographystyle{IEEEtran}
\bibliography{mybib}

\end{document}